\documentclass[prl,aps,amsmath]{revtex4}
\usepackage{color}
\newcommand{\half}{\frac{1}{\sqrt{2}}}
\newcommand{\dm}{\textrm{dim}}

\begin{document}

\title{No-signaling from Gleason non-contextuality and the Tensor Product
  Structure}

\author{Himanshu Sharma}
\author{R Srikanth}
\email{srik@poornaprajna.org}
\affiliation{Poornaprajna Institute of Scientific Research, Sadashivnagar,
Bengaluru- 560080, India.}
\affiliation{Raman Research Institute, Bengaluru- 560080, India.}

\begin{abstract}
The no-signaling principle in quantum mechanics is shown to be a 
consequence of Gleason non-contextuality and the tensor product
structure. 
\end{abstract}

\maketitle

\section{Introduction}

Gleason's  theorem \cite{gle}  asserts  that for  a  Hilbert space  of
dimension  3  or  greater,   the  only  possible  probability  measure
$\mu(\alpha)$ associated with a particular linear subspace $\alpha$ of
a Hilbert space will  have the form Tr$(\hat{\Pi}_\alpha \rho)$, where
$\hat{\Pi}_\alpha$ is the projector to the subspace, and $\rho$ is the
density matrix  for the  system.  The premise  needed for  proving the
theorem, apart from certain continuity requirements, is the assumption
of  non-contextuality:  the  probability measure,  $\sum_j  \mu(e_j)$,
associated with  a Hilbert subspace,  is independent of the  choice of
basis (context), $\{e_j\}$.  The no-signaling theorem as applicable to
a composite system, essentially  asserts that in a bipartite composite
system consisting of  parts $A$ and $B$, the  reduced density operator
$\rho_A$ is unaffected by local operations in $B$ \cite{srik}.

For our purpose, it will be  useful to restate these two principles as
follows.     Given   a    system   $S$,    we   say    that    it   is
\textit{tensor-product}   partitioned  into   sectors  $J_i$   if  the
respective Hilbert spaces satisfy
\begin{equation} {\cal H}_S = \bigotimes_i {\cal J}_i,
\label{eq:part}
\end{equation}
with  $\dm({\cal H}_S)  =  \Pi_i \dm({\cal  J}_i)$.  The  no-signaling
theorem  asserts  that  the  marginal probability  distribution  $p_i$
associated  with sector $J_i$  is unaffected  by local  operations (in
particular, measurement in some basis) in other sectors $J_{i^\prime}$
(i.e.,   operations  of   the   form  $I_{i^\prime   \ne  i}   \otimes
O_{i^\prime}$).   It is  customary to  think of  the sectors  $J_i$ as
being  spatially  separated in  order  to  make  the term  `signaling'
meaningful,  though  spatiality is  not  essential  to the  formalism.
No-signaling applies  at the single-particle level  also, as discussed
below.

We say  that it is \textit{tensor-sum} partitioned  into sectors $K_j$
if the respective Hilbert spaces satisfy
\begin{equation} {\cal H}_S = \bigoplus_j {\cal K}_j
\label{eq:sector}
\end{equation}
with  $\dm({\cal  H}_S)  =   \sum_j  \dm({\cal  K}_j)$.   The  Gleason
non-contextuality  assumption  asserts  that the  probability  measure
$q_j$ associated  with sector $K_j$ is unaffected  by local operations
in that sector, i.e., rotations  of the basis vectors such that ${\cal
  K}_j$ is  an invariant subspace  of the operations. In  other words,
the choice of  measurement basis in that sector,  or by extension, the
projectors used to complete the  full basis in the other sectors, does
not alter the probability measure associated with $K_j$.

Expressed  thus,  the   result  we  wish  to  prove,   which  is  that
no-signaling in  a multi-partite system  is a manifestation  of single
system   non-contextuality  in   a  tensor   product   setting,  seems
tantalizingly plausible.  Stated differently, we wish to show that the
mutual   independence   of    marginal   probabilities   under   local
transformations in  distinct sectors across  a tensor \textit{product}
cut reduce to the independence of probability measures across a tensor
\textit{sum} cut.

\section{Single-particle case}

Suppose we  are given the orthogonal  states of a  qutrit, whose state
space  ${\cal H}$  is spanned  by the  basis  $\{|0\rangle, |1\rangle,
|2\rangle\}$.  Alice  and Bob  are two spatially  separated observers.
Incoming qutrits prepared in  the state $|\psi\rangle = \sum_j \beta_j
|j\rangle$ ($j=0,1,2$ and  $\sum_j |\beta_j|^2=1$) are first subjected
to a \textit{non-maximal test} corresponding to the measurement of the
degenerate observable $X = a |0\rangle\langle0| + b(|1\rangle\langle1|
+      |2\rangle\langle2|)      =      a     |0\rangle\langle0|      +
b(|{+}\rangle\langle{+}|     +     |{-}\rangle\langle{-}|)$,     where
$|\pm\rangle  \equiv \half(|1\rangle  \pm |2\rangle)$  and $a,  b$ are
real numbers.  Alice is located  in station $A$ where she receives the
qutrit if measurement of $X$ returns $a$.  Otherwise the particle goes
to Bob, who  measures it either in the  basis $Y_1 \equiv \{|1\rangle,
|2\rangle\}$ or in the basis $Y_2 \equiv \{|\pm\rangle\}$.

To   show   that  non-contextuality   entails   no-signaling  at   the
single-particle level, assume that there is a nonlocal signal from Bob
to  Alice, depending  on whether  he measures  in the  basis  $Y_1$ or
$Y_2$.   An  instance  of   signaling  implies  that  the  probability
prob($|0\rangle$)  that  Alice finds  the  particle  depends on  Bob's
choice of basis in his subspace \cite{peres}:
$$\textrm{Prob}(|0\rangle|Y_1)  \ne  \textrm{Prob}  (|0\rangle|Y_2).$$
But  this  means that  the  probability  measure  associated with  the
subspace span$\{|1\rangle,|2\rangle\}$ depends on the choice of basis,
and  therefore   provides  a   mechanism  by  which   the  probability
Prob$(|0\rangle)$   is  contextual,   and  we   obtain   our  required
result. The converse is not  true, however. In this example, if ${\cal
  H}$  corresponded to  an internal  degree  of freedom,  so that  the
states  $|j\rangle$  ($j=0,1,2$)  are  not spatially  separated,  then
contextuality would not lead to signaling.

Non-contextuality   also    implies   the   no-disturbance   principle
\cite{rama}, which  may be thought of as  non-contextuality applied to
whole observables  rather than individual events  or projectors.  For,
by construction,
\begin{equation}
[X,Y_j] = 0;~ [Y_1^+, Y_2^+] \ne 0,
\label{eq:nodist}
\end{equation}
where  $Y_j^+  =  \{|0\rangle\}   \cup  Y_j$.   Thus,  if  there  were
disturbance, namely  the $X$ marginals  satisfy $\sum_i P(X=x,Y_1^+=i)
\ne \sum_j P(X=x,Y_2^+=j)$, this would constitute context dependence.

\section{Multi-particle case}

To  prove  that non-contextuality  implies  no-signaling  in a  tensor
product setting, we consider a bi-partite system, given by the Hilbert
space ${\cal H}_S  \equiv {\cal H}_A \otimes {\cal  H}_B$.  Our result
can straightforwardly be generalized to larger systems.  By assumption
of  tensor product  structure,  a  local operation  on  $A$'s side  (a
unitary operation  followed by projective  measurement or a  POVM) has
the form  $\omega \equiv {\cal E}_A  \otimes I_B$, where  $I_B$ is the
identity  operation in  ${\cal H}_B$.   And  similarly $\omega^\prime$
corresponding  to another local  operation on  $A$'s side.   Given any
tensor-sum partition:
\begin{equation}
{\cal  H}_S   =  \bigoplus_j  {\cal  K}_j   \equiv  \bigoplus_j  {\cal
  H}_A\otimes {\cal B}_j,
\label{eq:x}
\end{equation}
where ${\cal H}_B = \bigoplus_j {\cal B}_j$, and let $\delta_j$ denote
the set  of dimensions that  span ${\cal B}_j$. Each  partition ${\cal
  K}_j$   is  an   invariant   subspace  under   $\omega$  and   under
$\omega^\prime$.  By non-contextuality,  the probability measure $\mu$
associated with  any given ${\cal  K}_j$ should be independent  of the
application of $\omega$ or $\omega^\prime$: i.e.,
\begin{equation}
\mu({\cal  K}_j|{\cal  E}) =  \mu({\cal  K}_j|{\cal E}^\prime)  \equiv
\mu({\cal K}_j).
\label{eq:nocon}
\end{equation}  
Now, 
\begin{subequations}
\begin{eqnarray}
\mu({\cal  K}_j|{\cal E})  &\equiv& \sum_a\sum_{k \in \delta_j}
\textrm{Prob}(A=a,B=k) = \textrm{Prob}_B(j|{\cal E}) \\ 
\mu({\cal  K}_j|{\cal  E^\prime}) &\equiv&
\sum_{a^\prime}\sum_{k \in \delta_j}  \textrm{Prob}(A^\prime=a^\prime, B=k)
= \textrm{Prob}_B(j|{\cal E}^\prime),
\end{eqnarray}
\label{eq:mu}
\end{subequations}
where $\textrm{Prob}_B(j|\xi)$  is the  probability for Bob  to obtain
outcome $j$ in the $\xi$-context  $(\xi = {\cal E}, {\cal E}^\prime)$.
If signaling  were possible, it  would mean that  there is a  $j$ such
that
\begin{equation}
\textrm{Prob}_B(j|{\cal E}) \ne \textrm{Prob}_B(j|{\cal E}^\prime),
\end{equation}
which in view of Eq. (\ref{eq:mu}), implies
\begin{equation}
\mu({\cal K}_j|{\cal E}) \ne \mu({\cal K}_j|{\cal E^\prime}).
\end{equation}
Together  with  Eq.  (\ref{eq:nocon}),  this  implies  a violation  of
non-contextuality  in sector  ${\cal  K}_j$.  This  proves our  stated
result, which, as  it happens, connects the Born  rule to no-signaling
via  Gleason's theorem.   

An instance of  contexuality, on the other hand,  does not necessarily
lead  to  signaling  across  spatially  separated  sectors  (as  noted
earlier) or  across a tensor product  cut.  As example  of the latter:
contextuality in a system with prime-numbered dimensionality cannot be
represented  as  a signaling  across  two  tensor  product sectors  of
non-trivial dimensionality.  We may  regard signaling as the avatar of
contextuality in  a spatial situation where some  events correspond to
geographically  separated  locations.   Non-contextuality  then  is  a
stronger  condition  than   no-signaling  assuming  a  tensor  product
structure.

A  proof   of  no-signaling  also   obtains  as  a  special   case  of
no-disturbance, where $X$ and either $Y_j^+$ in Eq.  (\ref{eq:nodist})
are assumed  to correspond to two different  particles (tensor product
sectors). However, when  $X$ and $Y_j^+$ belong to  the same particle,
as we  saw, $X$  must be  non-maximal or degenerate  (at least  in the
subspace where  $Y_1^+$ and  $Y_2^+$ fail to  commute).  On  the other
hand,  no such  restriction appears  when $X$  and $Y_j^+$  pertain to
distinct particles  or degrees  of freedom or  tensor-product sectors.
This extra  difference between the  single-particle and multi-particle
situation  does not  appear in  our proof,  where the  more elementary
events rather than  observables are the primary objects,  and thus the
reduction is unconditional.

\section{Discussion and Conclusions}

We may  regard non-contextuality as a more  fundamental principle than
no-signaling  for several  reasons.  It  pertains to  a  single system
rather than a composite system.  It enables unifying no-signaling into
a  single  stronger  no-go   principle,  as  noted  above.   Moreover,
no-signaling in  our present sense arises  in non-relativistic quantum
mechanics.  For it to be aligned with \textit{relativisitic} causality
would be  an odd conspiracy,  that would need further  explanation. No
such difficulty  arises when we  regard non-contextuality as  the more
fundamental principle, with  no-signaling an `innocent' consequence of
imposing it on a tensor-product structured space.

While no-signaling is  no doubt a useful thumb-rule  in deriving other
results  (e.g.,  as in  Ref.  \cite{hom,clon}),  we  believe that  our
observation  would   be  of   interest  in  axiomatic   studies  where
no-signaling is treated as a primary postulate \cite{gis,dan}.  It can
also help  clarify how, if potential violations  of no-signaling arise
in a  more general theory than  quantum mechanics, they  may reduce or
relate  to  single-particle effects.   In  Ref.   \cite{srik0}, it  is
suggested that quantum optics is  testably such a more general theory,
with  peculiarities  introduced  because  of the  lower-boundeness  of
energy imposed by the vacuum state.

Gleason  non-contextuality is  intimately  related to,  and yet  quite
distinct from, Kochen-Specker  (KS) contextuality \cite{ks,winter}, on
which we report  elsewhere \cite{lotus}.  It is the  latter that makes
the  former  surprising.   To  use existing  terminology  (as  usually
applied  to  quantum  nonlocality  in  the  multi-partite  situation),
Gleason  non-contextuality refers  to  \textit{parameter independence}
\cite{desp}   or   signal   locality,   while  KS   contextuality   to
\textit{outcome  dependence}  \cite{desp}  or  violation  of  Einstein
locality.  Parameter independence by  itself would demand no more than
`garden variety' classicality: as  for example, with vector components
whose magnitude  is given by the 2-norm,  such as the length  of a 3-D
object or  intensity of  the electric field  along a  given direction.
Outcome dependence  by itself would  demand a disturbance  produced by
measurement or  (in a spatial setting)  superluminal classicality.  It
is putting the two together  that requires the subtle richness that is
quantum   contextuality   or  quantum   nonlocality,   based  on   the
non-commutative   structure  of  quantum   mechanics,  or   any  other
generalized non-signaling probability distributions \cite{boxy,lla}.

It  has   recently  been  shown  that  local   quantum  mechanics  and
no-signaling  imply quantum  correlations  \cite{wehner}.  Our  result
shows  that  assuming  tensor  product structure,  no-signaling  is  a
consequence of an  aspect of local quantum mechanics,  thus making the
case  for quantum  correlations stronger.   As one  consequence, local
quantum mechanics with super-quantum correlations entails superluminal
signaling \cite{arv}.

\end{document}